\documentstyle[11pt]{article}
\title{{\bf The myth of the down converted photon}}
\author{Trevor~W.~Marshall\\
Department of Mathematics, University of Manchester\\
Manchester M13 9PL, U.K.}
\begin{document}
\maketitle
\begin{abstract}
Parametric
down conversion (PDC)
is widely interpreted in terms of photons, but,
even among supporters of this interpretation,
many properties of the photon pairs have been
described as ``mind-boggling" and even ``absurd".
In this article we argue that a classical description
of the light field, taking account of its vacuum
fluctuations, leads us to a consistent and rational
description of all PDC phenomena. ``Nonlocality"
in quantum optics is simply an artifact of the
Photon Concept. We also predict a new phenomenon,
namely the appearance of a second, or satellite
PDC rainbow.
(This article will appear in the Proceedings of
the Second Vigier Conference held in York University,
Canada in August 1997. A somewhat more formal version
has been submitted to Phys. Rev. Letters, and
may be found at http:\slash\slash xxx.lanl.gov\slash abs
\slash quant-ph\slash
9711029.)
\end{abstract}
\section{Introduction}
In an article for the last conference in this series\cite{vig}
we gave a description of the Parametric Down Conversion (PDC)
process based on the real vacuum
electromagnetic-field fluctuations. We indicated that there
was a serious unsolved problem, in that detectors must
somehow subtract away these fluctuations; such a
mechanism must come into play in order to explain the
very low dark rates actually observed. We have since
published a series of
articles\cite{pdc1,pdc2,pdc3,pdc4}
in which a great
variety of PDC phenomena have been analyzed using
this description. Since we have
been able to establish a formal parallel, through
the Wigner representation, between the new
(or rather the old!) field description and the
presently dominant Photon Theory, it is
clear that, {\it once the reality of the zeropoint
field has been accepted, there are no PDC
phenomena which require photons.}
Furthermore we have made considerable progress
on the subtraction problem\cite{pdc4}; all that
is needed to explain the low dark rates of detectors is
the recognition of their extremely large time
windows (5ns is a very large number of light oscillations).

The approach of the above series of articles
was a kind of compromise between the standard
nonlocal theory of Quantum Optics, where the
interaction of the various field modes is
represented by a hamiltonian, and a fully
maxwellian theory, which would be both
local and causal. In this latter case the
nonlinear crystal would be represented as
a spatially localized current distribution,
modified of course by the incoming
electromagnetic field; the outgoing field
would then be expressed as the retarded
field radiated by this distribution.
A preliminary attempt at such a theory was made\cite{magic},
using first-order perturbation theory.
However, we showed, in the above series of articles, that a calculation
of the relevant counting rates, to lowest order, requires us to find the
{\it second}-order perturbation corrections to the Wigner density,
and the close formal parallel between these two theories
means that the same considerations will apply to
the maxwellian theory.

\section{What is PDC?}
It is necessary to pose this question, because,
depending on the answer given, PDC may be
described as either a local or a nonlocal
phenomenon.

An example of the modern, nonlocal description
is provided by Greenberger, Horne and Zeilinger\cite{ghz}.
A nonlinear crystal, pumped by a laser at frequency
$\omega_0$, produces conjugate pairs of signals,
of frequency $\omega$ and $\omega_0-\omega$ (see Fig.1).
Since light is supposed to consist
of photons, this means that an incoming laser
photon ``down converts" into a pair of lower-energy
photons. Naturally, since we know that $E=\hbar\omega$,
that means energy is conserved in the PDC process,
which must be very comforting. However, the above
authors themselves refer to the PDC photon-counting
statistics as ``mind-boggling", and a more recent
commentary\cite{zeil} even uses the term ``absurd".
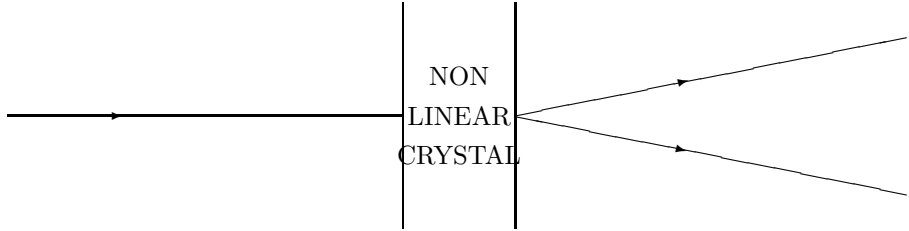
\begin{figure}[htb]
\unitlength=0.75mm
\linethickness{0.4pt}
\begin{picture}(159.33,44.67)
\put(70.00,44.67){\line(0,-1){40.00}}
\put(90.00,4.67){\line(0,1){40.00}}
\thicklines
\put(0.00,24.67){\line(1,0){70.00}}
\thinlines
\put(90.00,24.67){\line(5,1){69.33}}
\put(159.33,10.67){\line(-5,1){69.33}}
\put(80.00,24.67){\makebox(0,0)[cc]{{\small LINEAR}}}
\put(80.00,31.67){\makebox(0,0)[cc]{{\small NON}}}
\put(80.00,17.67){\makebox(0,0)[cc]{{\small CRYSTAL}}}
\put(12.67,24.67){\vector(1,0){7.67}}
\put(119.33,18.67){\vector(4,-1){1.00}}
\put(119.67,30.67){\vector(4,1){1.00}}
\end{picture}

\caption{PDC - the modern version. A laser photon
down converts into a conjugate pair of PDC photons
with conservation of energy.}
\end{figure}

There is an older description, which I suggest is
more correct than the modern one. It had only a short life.
Nonlinear optics was born in the late 1950s, with the
invention of the laser. Up to about 1965, when
Quantum Optics was born, the PDC process would
have been depicted\cite{saleh,yariv} by Fig.2; an incoming wave
of frequency $\omega$ is down converted, by
the pumped crystal, into an
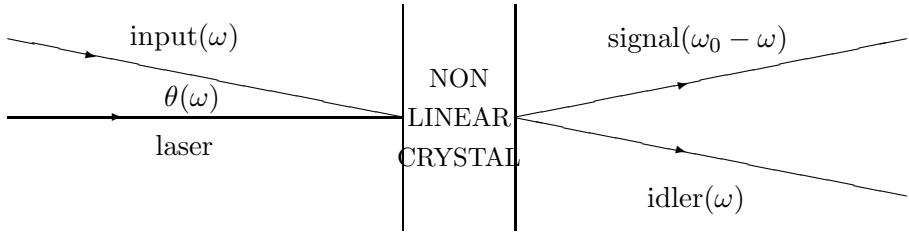
\begin{figure}[htb]
\unitlength=0.75mm
\linethickness{0.4pt}
\begin{picture}(159.33,44.67)
\put(70.00,44.67){\line(0,-1){40.00}}
\put(90.00,4.67){\line(0,1){40.00}}
\thicklines
\put(0.00,24.67){\line(1,0){70.00}}
\thinlines
\put(70.00,24.67){\line(-5,1){70.00}}
\put(90.00,24.67){\line(5,1){69.33}}
\put(159.33,10.67){\line(-5,1){69.33}}
\put(31.33,19.67){\makebox(0,0)[cc]{laser}}
\put(31.33,38.67){\makebox(0,0)[cc]{input$(\omega)$}}
\put(122.33,38.34){\makebox(0,0)[cc]{signal$(\omega_0-\omega)$}}
\put(122.00,10.00){\makebox(0,0)[cc]{idler$(\omega)$}}
\put(80.00,24.67){\makebox(0,0)[cc]{{\small LINEAR}}}
\put(80.00,31.67){\makebox(0,0)[cc]{{\small NON}}}
\put(80.00,17.67){\makebox(0,0)[cc]{{\small CRYSTAL}}}
\put(12.67,24.67){\vector(1,0){7.67}}
\put(14.67,35.67){\vector(4,-1){1.00}}
\put(119.33,18.67){\vector(4,-1){1.00}}
\put(119.67,30.34){\vector(4,1){1.00}}
\put(32.67,28.00){\makebox(0,0)[cc]{$\theta(\omega)$}}
\end{picture}

\caption{PDC - the ancient version. When a wave of
frequency $\omega$ is incident, at a certain angle
$\theta(\omega)$, on a nonlinear crystal
pumped at frequency $\omega_0$, a signal
of frequency $\omega_0-\omega$ is emitted
in a certain conjugate direction. The modified
input wave is called the idler.}
\end{figure}
outgoing signal
of frequency $\omega_0-\omega$. The explanation
of the frequency relationships lies in the
multiplication, by the nonlinear crystal,
of the two input amplitudes; we have no need of $\hbar$!

This process persists when the intensity
of the input is reduced to zero, because
all modes of the light field are still
present in the vacuum, and the nonlinear crystal
modifies vacuum modes in exactly the same way
as it modifies input modes supplied by an
experimenter.
What we see  emerging from the crystal is
the familiar PDC rainbow.
This is because the angle of incidence
$\theta$, at which PDC
occurs, is different for different frequencies
on account of the variation of refractive
index with frequency.

We depict the
process of PDC from the vacuum in Fig.3,
but note that this figure shows only two
conjugate modes of the light field; a complete
picture would show all frequencies participating
in conjugate pairs, with varying angles
of incidence. In contrast with Fig.2, where we
showed only the one relevant input, we must now
take account also of the conjugate input mode
of the zeropoint, since the first mode itself has
only the zeropoint amplitude.

The zeropoint inputs, denoted by interrupted lines in
Fig.3, do not activate photodetectors, because the
threshold of these devices is set precisely at the
level of the zeropoint intensity, as discussed in Ref.\cite{pdc4}.
\begin{figure}[htb]
\unitlength=0.75mm
\linethickness{0.4pt}
\begin{picture}(159.33,44.67)
\put(70.00,44.67){\line(0,-1){40.00}}
\put(90.00,4.67){\line(0,1){40.00}}
\thicklines
\put(0.00,24.67){\line(1,0){70.00}}
\thinlines
\put(31.67,28.34){\makebox(0,0)[cc]{$\theta(\omega)$}}
\put(90.00,24.67){\line(5,1){69.33}}
\put(159.33,10.67){\line(-5,1){69.33}}
\put(11.33,19.67){\makebox(0,0)[cc]{laser}}
\put(31.33,38.67){\makebox(0,0)[cc]{input($\omega$)}}
\put(122.33,38.34){\makebox(0,0)[cc]{+idler($\omega_0-\omega$)}}
\put(122.00,10.00){\makebox(0,0)[cc]{idler($\omega$)}}
\put(80.00,24.67){\makebox(0,0)[cc]{{\small LINEAR}}}
\put(80.00,31.67){\makebox(0,0)[cc]{{\small NON}}}
\put(80.00,17.67){\makebox(0,0)[cc]{{\small CRYSTAL}}}
\put(12.67,24.67){\vector(1,0){7.67}}
\put(14.67,36.00){\vector(4,-1){1.00}}
\put(119.33,18.67){\vector(4,-1){1.00}}
\put(119.67,30.67){\vector(4,1){1.00}}
\put(70.00,24.67){\line(-5,1){42.33}}
\put(15.67,35.67){\line(-5,1){15.67}}
\put(0.00,38.13){\line(0,0){0.00}}
\put(31.33,44.00){\makebox(0,0)[cc]{zeropoint}}
\put(122.67,43.67){\makebox(0,0)[cc]{signal($\omega_0-\omega$)}}
\put(122.00,5.33){\makebox(0,0)[cc]{+signal($\omega$)}}
\put(0.00,10.67){\vector(4,1){16.00}}
\put(70.00,24.67){\line(-5,-1){39.67}}
\put(31.33,13.33){\makebox(0,0)[cc]{zeropoint}}
\put(31.33,9.00){\makebox(0,0)[cc]{input($\omega_0-\omega$)}}
\put(33.33,21.67){\makebox(0,0)[cc]{$\theta(\omega_0-\omega)$}}
\end{picture}

\caption{PDC from the vacuum. Both of
the outgoing signals are above zeropoint intensity, and
hence give photomultiplier counts.}
\end{figure}
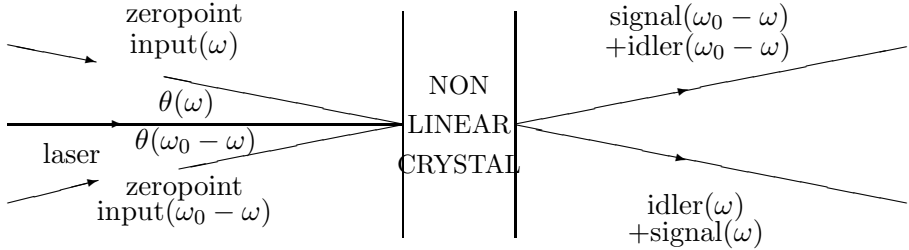
However,
the two idlers have intensities above
that of their corresponding inputs.
Also there is no
coherence between a signal and an idler of the same
frequency, so  their intensities
are additive in both channels. Hence there are photoelectron
counts in both of the outgoing channels of Fig.3.

The question we have posed in this section could
be rephrased as ``What is it that is down converted?".
According to the thinking behind Fig.1, the laser
photons are down converted, whereas according to
Fig.3 it is the zeropoint modes; they undergo both
down conversion, to give signals, and amplification,
to give idlers.

\section{Photon production rates in PDC}
There is a small, but important difference between
the maxwellian theory and the theory outlined in
our Wigner series\cite{pdc1,pdc2,pdc3,pdc4}, though
both of them could be said to be based on Fig.3.
The Wigner series gave us the undulatory version
of quantum optics, but its starting point is a
hamiltonian which takes the creation of photon
pairs as axiomatic. The maxwellian theory, whose
details are given elsewhere\cite{puc1}, starts from
a nonlinear expression for the induced current
and deduces a coupling between the field modes.
This coupling is very similar, but not identical,
to that deduced from the Wigner-based theory.
As we have emphasized, there are no photons
in the maxwellian theory, but if we
translate the intensities of the outgoing
signals in Fig.3 into photon terms, we obtain the result
\begin{equation}
\frac{n_i(\omega)+n_s(\omega)}{n_i(\omega_0-\omega)+n_s(\omega_0-\omega)}=
\frac{\cos[\theta(\omega_0-\omega)]}
{\cos[\theta(\omega)]}\;.   \label{pdcint}
\end{equation}

So we conclude that {\it the photon rate in
a given channel is inversely proportional to
the cosine of the rainbow angle}. In the Photon
Theory, the above ratio is one.

There seems little chance of finding out
directly which of these theories is correct;
the difference between the two ratios is small,
since the rainbow angles are typically around
10 degrees, and it is not possible to
measure at all accurately the efficiency
of light detectors as a function of
frequency. It is true that some of the
experiments we have analysed, using the
standard theory, in Refs.\cite{pdc1,pdc2,pdc3,pdc4},
have slightly different results in the present
theory, for example the fringe visibility in
the experiment of Zou, Wang and Mandel\cite{zwm}.
Some details will be published shortly, but we
can say that an experimental discrimination will
be very difficult.

\section{Parametric up conversion from the vacuum}

There is, however, at least one prediction
of the new theory which differs dramatically
from the standard theory. An incident wave
of frequency $\omega$, as well as being
down converted by the pump to give
a PDC signal of frequency $\omega_0-\omega$,
may also be {\it up converted to give a
PUC signal} of frequency $\omega_0+\omega$.
We depict this phenomenon, which is well
known\cite{saleh,yariv} in classical nonlinear optics,
in Fig.4.
\begin{figure}[htb]
\unitlength=0.75mm
\linethickness{0.4pt}
\begin{picture}(159.33,86.80)
\put(70.00,85.00){\line(0,-1){80.00}}
\put(90.00,5.00){\line(0,1){80.00}}
\thicklines
\put(0.00,45.00){\line(1,0){70.00}}
\thinlines
\put(80.00,45.00){\makebox(0,0)[cc]{{\small LINEAR}}}
\put(80.00,52.00){\makebox(0,0)[cc]{{\small NON}}}
\put(80.00,38.00){\makebox(0,0)[cc]{{\small CRYSTAL}}}
\put(12.67,45.00){\vector(1,0){7.67}}
\put(70.00,45.00){\line(-5,3){69.67}}
\put(90.00,45.00){\line(5,-3){69.33}}
\put(159.33,3.40){\line(0,0){0.00}}
\put(90.00,45.00){\line(6,-1){69.33}}
\put(49.00,51.00){\makebox(0,0)[cc]{$\theta_u(\omega)$}}
\put(32.33,38.67){\makebox(0,0)[cc]{laser}}
\put(33.67,73.67){\makebox(0,0)[cc]{input($\omega$)}}
\put(131.67,43.33){\makebox(0,0)[cc]{signal($\omega_0+\omega$)}}
\put(115.67,21.00){\makebox(0,0)[cc]{idler($\omega$)}}
\put(120.00,27.00){\vector(3,-2){1.00}}
\put(121.67,39.67){\vector(4,-1){1.00}}
\put(27.67,70.33){\vector(3,-2){1.00}}
\end{picture}

\caption{PUC. In contrast with PDC
the output signal has its
transverse component in the same direction as that of
the idler.}
\end{figure}
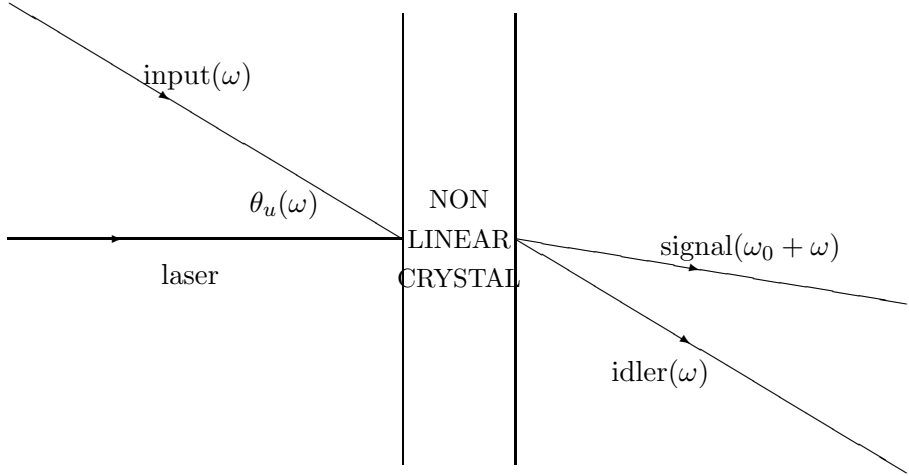
Note that the angle of incidence, $\theta_u(\omega)$,
at which PUC occurs is quite different from the PDC
angle, which in Fig.2 was denoted simply $\theta(\omega)$,
but which we should now call $\theta_d(\omega)$.

Now, following the same argument which led us from
Fig.2 to Fig.3, we predict the phenomenon of
PUC from the Vacuum, which we depict in Fig.5.

When we come to calculate the intensity of the PUC rainbow,
there is an important difference from the PDC situation, because
we find that the idler intensities are now less than the
input zeropoint intensities. The signal intensities in
both channels almost, but not quite, cancel this shortfall,
so that the PUC intensities are only about
3 per cent of the PDC intensities, which may explain
why nobody has yet observed them. Also, note that there
is a detectable signal only in the lower-frequency
channel, because the relation corresponding to eq.(\ref{pdcint})
is
\begin{equation}
\frac{n_i(\omega)+n_s(\omega)}{n_i(\omega_0+\omega)+n_s(\omega_0+\omega)}=
-\frac{\cos[\theta_u(\omega_0+\omega)]}
{\cos[\theta_u(\omega)]}\;,
\end{equation}
which means that in one of the channels (actually the upper-frequency
one), the total output intensity is less than the zeropoint,
so nothing will be detected in this channel.
My prediction therefore is that, as well as the main
PDC rainbow $\theta_d(\omega)$, {\it there is also a satellite
rainbow}, whose intensity is about 3 percent of the
main one, at $\theta_u(\omega)$.
An approximate calculation\cite{puc1} shows that
$\theta_u(\omega)$ is about 2.5 times $\theta_d(\omega)$.
\begin{figure}[htb]
\unitlength=0.75mm
\linethickness{0.4pt}
\begin{picture}(159.33,85.00)
\put(70.00,85.00){\line(0,-1){80.00}}
\put(90.00,5.00){\line(0,1){80.00}}
\thicklines
\put(0.00,45.00){\line(1,0){70.00}}
\thinlines
\put(80.00,45.00){\makebox(0,0)[cc]{{\small LINEAR}}}
\put(80.00,52.00){\makebox(0,0)[cc]{{\small NON}}}
\put(80.00,38.00){\makebox(0,0)[cc]{{\small CRYSTAL}}}
\put(12.67,45.00){\vector(1,0){7.67}}
\put(90.00,45.00){\line(5,-3){69.33}}
\put(159.33,3.40){\line(0,0){0.00}}
\put(32.33,38.67){\makebox(0,0)[cc]{laser}}
\put(33.67,73.67){\makebox(0,0)[cc]{input($\omega$)}}
\put(131.67,43.33){\makebox(0,0)[cc]{+idler($\omega_0+\omega$)}}
\put(115.67,21.00){\makebox(0,0)[cc]{idler($\omega$)}}
\put(120.00,27.00){\vector(3,-2){1.00}}
\put(121.67,39.67){\vector(4,-1){1.00}}
\put(27.67,70.00){\vector(3,-2){1.00}}
\put(33.33,79.67){\makebox(0,0)[cc]{zeropoint}}
\put(70.00,45.00){\line(-6,1){30.67}}
\put(18.00,54.00){\line(-6,1){18.00}}
\put(13.00,55.00){\vector(4,-1){1.00}}
\put(25.00,56.33){\makebox(0,0)[cc]{input($\omega_0+\omega$)}}
\put(25.00,61.33){\makebox(0,0)[cc]{zeropoint}}
\put(131.33,50.00){\makebox(0,0)[cc]{signal($\omega_0+\omega$)}}
\put(115.67,14.67){\makebox(0,0)[cc]{+signal($\omega$)}}
\put(70.00,45.00){\line(-5,3){27.00}}
\put(28.67,69.33){\line(-5,3){22.33}}
\put(90.00,45.00){\line(6,-1){32.33}}
\put(137.33,36.33){\line(6,-1){21.33}}
\end{picture}

\caption{PUC from the vacuum.  Only one
of the outgoing signals is above the zeropoint intensity. The
other one, depicted by an interrupted line, is below
zeropoint intensity.}
\end{figure}
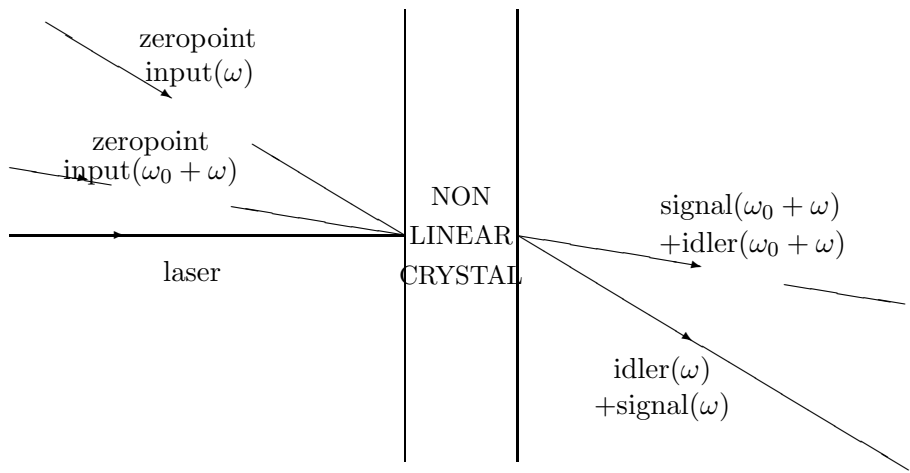

\newpage
\section{Conclusion}
Our contribution to the previous conference in this
series\cite{vig} was entitled ``The myth of the photon''.
The present article repeats this theme, but covers a
narrower range of phenomena. This is because the local
theory of nonlinear crystals is now very much more
complete than the corresponding theory for atoms. In
retrospect, the word ``obsolete'', which we used
in the previous article, for {\it all} photon theories,
was excessively triumphalist.
Of course one could argue that
they  became obsolete once their nonlocal
nature was revealed, that is a quarter of a century ago,
but there was nothing local on offer at that time.
The claim we made, maybe prematurely, was
based on having demonstrated, by the use of
certain model theories with very limited fields
of application, that local theories were, in
all cases {\it possible}. Now we have passed
into a new phase of the programme; we now have,
for a very wide and growing area of investigation,
{\it a well defined alternative theory which
makes certain new predictions}. If and when such
predictions are verified, I think that
down-converted photons, for example those
depicted in our Fig.1, will be
very definitely obsolete.

\noindent
{\bf Acknowledgement}

\noindent
I have had a lot of help with the ideas behind this article,
and also in developing the argument, from Emilio Santos.


\begin{thebibliography}{99}
\bibitem{vig}
T.~W. Marshall and E.~Santos,
{\it The myth of the photon} in
{\it The Present Status of the Quantum
  Theory of Light}, eds. S.~Jeffers et~al,
(Kluwer, Dordrecht, 1997) pages 67--77.
See also http:\slash\slash xxx.lanl.gov\slash abs
\slash quant-ph\slash 9711046.
\bibitem{pdc1} A. Casado, T. W. Marshall, and E. Santos,
{\it J. Opt. Soc. Am. B}, {\bf 14}, 494--502 (1997).
\bibitem{pdc2} A. Casado, A. Fernandez Rueda, T. W. Marshall,
R. Risco Delgado, and E. Santos, {\it Phys.Rev.A}, {\bf 55},
3879--3890 (1997).
\bibitem{pdc3} A. Casado,
A. Fernandez Rueda, T. W. Marshall, R. Risco Delgado,
and E. Santos,
{\it Phys.Rev.A}, {\bf 56}, 2477-2480 (1997)
\bibitem{pdc4}
A. Casado, T. W. Marshall and E. Santos, {\it J. Opt. Soc Am. B}
(awaiting publication).
See also http:\slash\slash xxx.lanl.gov\slash abs
\slash quant-ph\slash 9711042.
\bibitem{magic}
T. W. Marshall, {\it Magical Photon or Real Zeropoint?}
in {\it New Developments on Fundamental Problems in Quantum Physics},
eds.
M. Ferrero and A. van der Merwe (Kluwer, Dordrecht, 1997)
\bibitem{ghz}
D. M. Greenberger, M. A. Horne and A. Zeilinger,
{\it Phys. Today}, {\bf 46} No.8, 22 (1993)
\bibitem{zeil}
D. Bouwmeester and A. Zeilinger, {\it Nature} {\bf 388},
827-828 (1997)
\bibitem{saleh}
B.E.A.Saleh and M.C.Teich, {\it Fundamentals
of Photonics}, (John Wiley, New York, 1991) Chap. 19
\bibitem{yariv}
A. Yariv, {\it Quantum Electronics}
(John Wiley, New York, 1989) Chaps. 16 and 17
\bibitem{puc1}
T. W. Marshall, submitted to {\it Phys. Rev. A}
See also http:\slash\slash xxx.lanl.gov\slash abs
\slash quant-ph\slash 9711030
\bibitem{zwm}
L. J. Wang, X. Y. Zou and L. Mandel,
{\it Phys. Rev. A}, {\bf 44}, 4614
(1991)
\end{thebibliography}
\end{document}